\documentclass[prd,twocolumn,showpacs]{revtex4}
\usepackage{graphicx,color}
\usepackage{amsmath,amssymb,amsfonts}
\begin{document}

\title{On Detection of Black Hole Quasi-Normal Ringdowns: 
Detection Efficiency and Waveform Parameter Determination in Matched Filtering}

\author{Yoshiki Tsunesada${}^1$, Nobuyuki Kanda${}^2$, Hiroyuki Nakano${}^2$, 
Daisuke Tatsumi${}^1$, Masaki Ando${}^3$, Misao Sasaki${}^4$, Hideyuki Tagoshi${}^5$, 
and Hirotaka Takahashi${}^{5,6}$
}

\affiliation{${}^1$National Astronomical Observatory of Japan, Tokyo 181-8588, Japan}
\affiliation{${}^2$Department of Mathematics And Physics, Graduate School of Science, 
Osaka City University, Osaka 558-8585, Japan}
\affiliation{${}^3$Department of Physics, University of Tokyo, Tokyo 113-0033, Japan}
\affiliation{${}^4$Yukawa Institute for Theoretical Physics, Kyoto University, 
Kyoto 606-8502, Japan}
\affiliation{${}^5$Department of Earth and Space Science, Graduate School of Science, 
Osaka University, Osaka 560-0043, Japan}
\affiliation{${}^6$Graduate School of Science and Technology, 
Niigata University, Niigata 950-2181, Japan}

\begin{abstract}
Gravitational radiation from a slightly distorted black hole with ringdown waveform
is well understood in general relativity. It provides a probe for direct 
observation of black holes and determination of their physical parameters,
masses and angular momenta (Kerr parameters). For ringdown searches using 
data of gravitational wave detectors, matched filtering technique is useful.
In this paper, we describe studies on problems in matched filtering analysis
in realistic gravitational wave searches using observational data.
Above all, we focus on template constructions, matches or signal-to-noise 
ratios (SNRs), detection probabilities for 
Galactic events, and accuracies in evaluation of waveform parameters or black hole 
hairs. In template design for matched filtering, search parameter ranges and
template separations are determined by requirements from acceptable maximum loss
of SNRs, detection efficiencies, and computational costs.
In realistic searches using observational data, however, effects of non-stationary
noises cause decreases of SNRs, and increases of errors in waveform parameter
determinations. These problems will potentially arise in any matched filtering 
searches for any kind of waveforms. To investigate them, we have performed matched 
filtering analysis for artificial ringdown signals which are generated with 
Monte-Carlo technique and injected into the TAMA300 observational data. 
We employed an efficient  method to construct a bank of ringdown filters recently 
proposed by Nakano {\it et al.}, and use a template bank generated from a criterion 
such that losses of 
SNRs of any signals do not exceed $2 \%$. We found that this criterion is fulfilled
in ringdown searches using TAMA300 data, by examining distribution of SNRs
of simulated signals. It is also shown that with TAMA300 sensitivity,
the detection probability for Galactic ringdown events is about $50 \%$ for black 
holes of masses greater than $20 M_{\odot}$ with SNR $> 10$. The accuracies in 
waveform parameter estimations are found to be consistent with the template spacings,
and resolutions for black hole masses and the Kerr parameters
are evaluated as a few $\%$ and $\sim 40 \%$, respectively. They can be improved
up to $< 0.9 \%$ and $< 24 \%$ for events of ${\rm SNR} \ge 10$
by using fine-meshed template bank in the hierarchical search strategy. 
\end{abstract}

\maketitle

\section{Introduction}\label{secIntro}
With recent progress in the development of gravitational wave detectors 
as resonant bars \cite{BARS} and laser interferometers \cite{TAMA,LIGO,GEO},
the detectors' sensitivities and operational stabilities have 
significantly been improved.
Several groups have performed long-term observations which 
provided data of good quality suitable for various kind of 
astrophysical gravitational wave searches. These are for ``chirp'' signals from 
inspiralling compact neutron star binaries, burst-like signals from stellar 
core collapses in supernovae, stochastic gravitational wave background, 
and continuous waves from pulsars. Thus well organized studies on data 
analysis techniques are required for gravitational wave detection and 
gravitational wave astronomy in near future.
For some classes of gravitational wave sources, data analysis techniques have 
already been investigated extensively, in particular as for inspiral searches.
The matched filtering technique is considered as one of the best method for this, 
since the gravitational waveform is theoretically predicted in good accuracy.
Although it is rather difficult to exploit an optimal method for gravitational 
wave bursts with undefined waveforms,
several methods and filters have been proposed (e.g. \cite{burst1,burst2}).

Gravitational radiation from a perturbed black hole, so called {\it ringdown},
is one of the classes of known-waveform gravitational wave sources.
The black hole perturbation in the late stage of a black hole formation, initiated
by a coalescence of a binary system for example, can be 
described by quasi-normal modes (QNMs). Therefore gravitational radiation is 
expected as of damped sinusoidal waveform. This is represented in time domain by
\begin{equation}
 h(t) \propto \exp \left(- \frac{\pi f_c}{Q} (t - t_0) \right) \sin \left(2 \pi f_c (t - t_0) + \phi_0 \right) \label{eq:waveform}
\end{equation}
where $f_c, Q, \phi_0, t_0$ are the central frequency, the quality factor of the ringing, the initial phase, and the arrival time, respectively. The QNMs were studied extensively 
by using Regge-Wheeler-Zerilli or Teukolsky formalism
in 1970's 
and 1980's (e.g. \cite{Regge_1957td,Zerilli_1971wd,Teukolsky1,Teukolsky2}), 
and several methods have been proposed for 
numerical computation of them (\cite{Detweiler_1980gk,QNM1,QNM2,QNM3}, and \cite{Kokkotas} for review 
and references therein).
Leaver developed a numerically stable method for calculation of
QNMs with good accuracies \cite{Leaver}, and his results have been verified by 
many authors with independent schemes. For the least damped mode of the QNM
with $l = m = 2$ spin-$2$ spheroidal harmonics ${}_{-2}S_{22}$, 
Echeverria have given analytic expressions 
for $f_c$ and $Q$ as functions of the black hole mass $M$ and the non-dimensional 
angular momentum (the Kerr parameter) $a = \left[0, 1\right)$, as \cite{Echeverria}
\begin{eqnarray}
 f_c \; [{\rm kHz}] &\simeq& 32 \left( \frac{M}{M_{\odot}} \right)^{-1} \times \left[ 1 - \alpha (1 - a)^{\beta} \right] \label{eq:fc} \\
 Q &\simeq& 2 ( 1 - a )^{\gamma} \label{eq:Q}
\end{eqnarray}
where $\beta = 3/10, \gamma = -9/20$ and $\alpha = 63/100$. These expressions
give ringdown parameters $(f_c, Q)$ with accuracies of $\sim 5 \%$.
For black holes with masses of $10 \sim 200 M_{\odot}$, the ringdown 
frequencies fall in the observational band $100 {\rm Hz} \sim {\rm kHz}$ of 
ground-based laser interferometric detectors. Matched filtering technique is 
useful to find ringdown signals in outputs of gravitational wave detectors
since their waveforms are modeled in a simple form as equation 
(\ref{eq:waveform}) in terms of the parameters $(f_c, Q)$ which can be easily 
converted into the black hole parameters $(M, a)$. 

The black hole ringdowns are very interesting in gravitational wave 
astronomy from some aspects. First, ringdown detection will be a direct confirmation
of the existence of astrophysical black holes. Second, it gives an another 
``mass window'' for observation of astrophysical objects by detecting 
gravitational waves. For example, for black hole binaries with several tens 
of solar-masses, frequencies of gravitational radiation in the inspiralling 
phase are out of the observational band of contemporary laser interferometers, 
while they can be detected via ringdown waves. 
In spite of its significance, there are fewer publications on ringdown searches
compared to those for inspiral or burst searches. Creighton reported the results of
a single filter search for a black hole with a particular mass and spin using
data of the Caltech 40-m interferometer \cite{Creighton}. Arnaud et al. discussed
a general template construction method for matched filtering searches and its 
application to the ringdown case \cite{Arnaud}. 
Recently Nakano {\it et al.} 
proposed an efficient tiling method for ringdown filters \cite{Nakano,Nakano2}. 

The aim of this paper is to 
discuss problems in matched filtering analysis in realistic search using 
observational data. In matched filtering with a template bank, the template 
separation is taken so that losses of signal-to-noise-ratios (SNRs) of any 
signals do not exceed a pre-selected value ({\it minimal match}), $MM$. 
The choice of $MM$ determines detection efficiencies and waveform parameter 
resolutions. Some template construction methods for a given $MM$ have already been 
proposed as \cite{Arnaud}, \cite{Nakano,Nakano2}.
In a realistic search, however, non-stationary noises in detector outputs would 
reduce SNRs, so it is not clear whether the criterion $\ge MM$ is still fulfilled. 
Furthermore, the template which gives the maximum of the 
matches is not necessarily the {\it nearest} to the signal, thus
the noise effects would also appear in waveform parameter resolutions.
To investigate these problems, we have carried out matched filtering searches
for artificial ringdown signals embedded in TAMA300 data with Monte-Carlo technique.
The ringdown signals with randomly selected waveform parameters are injected 
into TAMA300 data by software, and filtered with the template bank. 
We employ the template spacing method recently proposed by Nakano 
{\it et al.} \cite{Nakano,Nakano2}, which provides an efficient tiling scheme to
cover the parameter space with a fewer templates keeping the criterion $\ge MM$.
For gravitational wave sources, we consider
black holes distributed in the Galaxy, since magnitudes of ringdown waves from 
distances of Galactic scales ($\sim 10 \, {\rm kpc}$) are expected enough large to 
be detected with contemporary interferometors as TAMA300 (see FIG. \ref{fig:sp}). 
From the simulation results, we evaluated SNRs and their losses,
detection efficiencies with given SNR thresholds, 
and the waveform parameter resolutions.
We also discuss possibility of black hole ``spectroscopy'' with gravitational wave
observations by using very fine-meshed template bank and hierarchical search 
technique.
The code used here is the same which
will be used in full-scale event searches using TAMA300 data. 
The information obtained in this work are 
important for astrophysical interpretation of search results
and to set an observational limit on the ringdown event rate.

The plan of the remaining part of the paper is as follows.
In Sec. II, we briefly describe the matched filtering technique
and the template construction method employed here.  The details
of our simulation studies are given in Sec. III,
and the results are shown in Sec. IV. Section V is devoted to discussion.
Finally, in Sec. VI, we summarize and conclude our work.

\section{Matched Filtering}\label{secMF}
\subsection{Template spacings}\label{subsec:template}
The matched filtering method is to compute a cross-correlation ({\it match})
between detector outputs and 
each of reference waveforms as a ``template'' characterized with waveform
parameter(s), $X^{\mu} = (X^1, X^2) = (f_c, Q)$ in ringdown case, 
to find signals possibly embedded in the detector outputs. The match between
a signal $s(t)$ and a template $h(t)$ with the $k$-th waveform parameter $X^{\mu}_k$
is expressed in Fourier domain as
\begin{eqnarray}
\rho(X^{\mu}_k) = \left< s | h(X^{\mu}_k) \right> \propto \int df \,
{\tilde{s}(f) \tilde{h}^*(f; X^{\mu}_k) \over S_n(|f|)}
\label{eq:inner0}
\end{eqnarray}
where $\tilde{s}(f)$ and $\tilde{h}(f)$ express the Fourier transforms of functions of time
$s(t)$ and $h(t)$, 
the asterisk $(\ast)$ denotes the complex conjugation, and $S_n(|f|)$ is the power spectrum 
of the detector noise.
In a practical search, the signal $s$ in equation (\ref{eq:inner0})
is replaced by the sum with detector noise $n$, namely $o = n + s$.
The template bank is produced so that for any signals
there exists at least one template which gives the match greater than 
the {\it minimal match}, as $\left< \hat{h}(X^{\mu})|\hat{h}(X^{\mu}_k)\right> \ge MM$ for
${}^\exists X^{\mu}_k$, where $\hat{h}$ is normalized as $\left<\hat{h}|\hat{h}\right>=1$. 
The choice of $MM$ is from
statistical or astrophysical requirements (a larger $MM$ reduces matches in 
average and 
hence detection efficiencies), and computational tolerance 
(a smaller $MM$ results in an increase of total number of templates).
Thus the key of template construction is how cover the parameter space with 
a fewer templates for a given $MM$.

A {\it mismatch} of two waveforms, defined as $1 - ({\rm match})$, can be understood
as a {\it distance} between them. Thus to obtain appropriate template spacings and
``tile'' the parameter space, 
it is required to calculate the {\it metric} of the space \cite{owen}.
As seen from equation (\ref{eq:inner0}), the metric of the parameter space is 
determined by the waveform $h$ and the power spectrum of detector noise $S_n(|f|)$. 
In our previous papers \cite{Nakano,Nakano2}, 
we discussed a tiling method in the frequency domain, 
assuming the detector noise is white, i.e., $S_n(|f|) = {\rm const.}$ 
in equation (\ref{eq:inner0}). 
In this case, the metric and the template tiling can be formulated analytically, hence
it is easy to program the template spacing scheme in a search code. 

In \cite{Nakano,Nakano2}, we also discussed the validity of a use of this template 
bank in the case of a detector's colored noise power spectrum.
As a model of realistic noise power spectrum, we used 
a fitting curve (without lines, structures, fluctuations) 
of the noise power spectrum of TAMA300 at Data Taking 8 in 2003. 
We prepare a template bank using the above spacing. 
Signals are generated so that their amplitudes are normalized to unity. 
Then, we perform the matched filtering 
and evaluate the maximum of the match for each signal. 
It is found that, in most cases ($99.12\%$ in our calculation), 
the losses in SNR are not greater than $2\%$ which is expected 
from the pre-assigned minimal match for the white noise case. 
This shows that the template spacing can be used even 
in the case of colored noise power spectrum. The mean of the match is $0.993$.
However, in the real data analysis, we must treat non-stationary 
noise fluctuations in detector's output. 
In the following, we discuss the simulated ringdown signals 
injected into TAMA300 data.

\subsection{Matched Filter Code}\label{subsec:mfCode}
The matched filtering code used here has developed at National Astronomical 
Observatory of Japan (NAOJ) and at Osaka City University (OCU). 
The TAMA300 data are split into data chunks of about $52 \, {\rm sec}$, and the matched 
filtering code are applied for each of them.
We use the moving-average method
to obtain the noise power spectrum of detector outputs $S_n(|f|)$ as a frequency 
weighting function in equation (\ref{eq:inner0}) in the calculation of 
cross-correlations 
\cite{Tatsumi}.
We define the parameter ranges for $f_c$ and $Q$ as
$1.0 \times 10^2 \leq f_c \leq 2.5 \times 10^3$ Hz and $2.0 \leq Q \leq 33.3$, 
and take the minimal match $MM = 0.98$.
The total number of templates ${\cal N}$ amounts to $628$.
To analyze a data chunk of length $52 \, {\rm sec}$, it takes about $130$ 
seconds with the $628$ filters using a single CPU (Intel Pentium-IV 
$2.5 \, {\rm GHz}$). The TAMA300 data have analyzed with sixteen-node PC clusters 
at NAOJ and OCU. The time required for a full-scale search using $1000$ hours data
is about one week.

\begin{figure}
\begin{center}
\includegraphics[scale=0.59]{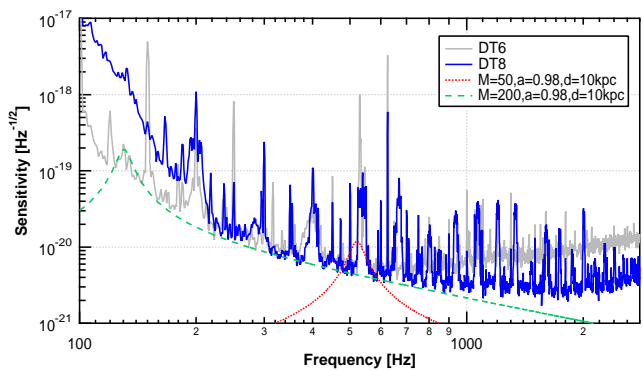}
\end{center}
\caption{TAMA300 sensitivity curves for DT6 (thin) and DT8 (thick).  Ringdown spectra are also shown for black holes of $M = 50 {\rm M}_{\odot}$ (dotted) and $M = 200 {\rm M}_{\odot}$ (dashed) with a fixed Kerr parameter $a = 0.98$ and a distance $d = 10 \, {\rm kpc}$. }
\label{fig:sp}
\end{figure}

In this work, we use the template bank generated so
that the minimal match is assured when the noise power spectrum is white
 i.e. $S_n(|f|) = {\rm const.}$ in the calculation of correlations of two 
waveforms by equation (\ref{eq:inner0}).
It must be noted that since the noise power spectrum of a real detector is not stationary,
the metric of the parameter space and appropriate template spacings also change in 
time. Nevertheless we apply the same
template bank to all the data chunks used here. This is supported by the fact that
ringdown waveforms have Fourier spectra of rather narrow band centered at $f_c$ 
(see FIG. \ref{fig:sp}), which is especially true for large $Q$ ringdowns, 
and the match given by equation (\ref{eq:inner0}) 
is expected to be less sensitive to whole shape of 
the noise power spectrum $S_n(|f|)$. This has been confirmed in \cite{Nakano2} 
with model calculations using the TAMA300 fitted noise curve as described above
(similar discussion is also found in \cite{Arnaud}).
Thus the question is if the template bank based on white noise power spectrum can still
be valid or effective enough when applied to real observational data that contain
time variation of detector's noise.

\section{Matched Filtering Simulation Using TAMA300 Data}\label{secMain}
\subsection{Monte-Carlo Simulation of Galactic Ringdown Events}
\begin{figure}
\begin{center}
\includegraphics[scale=0.7]{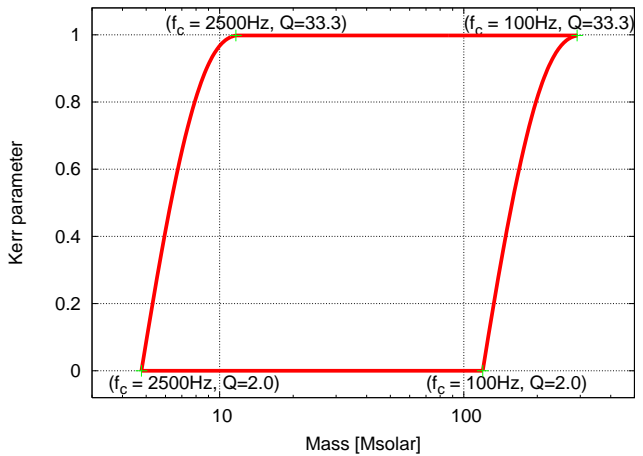}
\end{center}
\caption{Parameter range. We generate ringdown signals with parameters $f_c = 100 \sim 2500 {\rm Hz}, Q = 2 \sim 33.3$ uniformly chosen in $(f_c, Q)$ plane.}
\label{fig:parameter-range}
\end{figure}

\begin{figure}
\begin{center}
\includegraphics[scale=0.68]{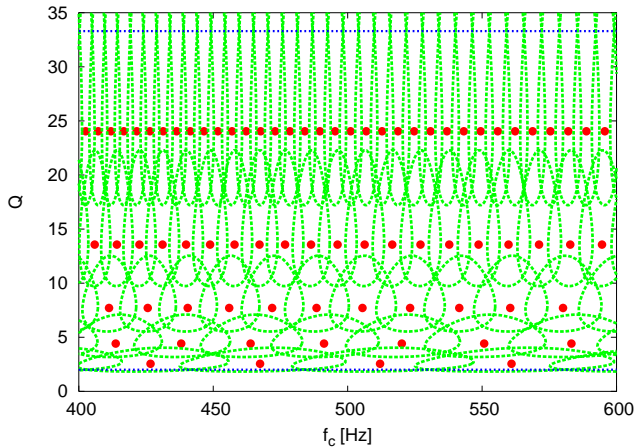}
\end{center}
\caption{A part of the template bank generated under the condition of minimal match 
$MM = 0.98$. The templates are placed along fixed $Q$ lines 
($Q = 2.6, 4.4, 7.7, 13.6, 24.0$) by using the tiling method proposed 
in \cite{Nakano}. The oval for each of the templates shows the tiled region
in which mismatches between the template and any waveforms 
do not exceed $2 \%$ in the case of white noise. 
There is no uncovered region in the parameter ranges $f_c = 100 \sim 2500 \, 
{\rm Hz}$ and $Q = 2 \sim 33.3$. The ratio of the total area
of the tiles to that of the parameter space covered is $1.57$.}
\label{fig:templates}
\end{figure}

In this section, we describe our simulations in detail. 
Randomly generated ringdown signals are injected into TAMA data, and filtered with the 
template bank. For gravitational wave sources, we consider Galactic black holes, since
TAMA300 has enough sensitivity to detect them thus they are suitable target for our search.
The spatial distribution of sources in the Galaxy is assumed as a form in
the Galacto-centric cylindrical coordinates $(R, z, \varphi)$,
\begin{eqnarray}
  {\rm d}N \propto \exp(-2 R^2/R_0^2) \exp(-|z|/h_z) R {\rm d}R {\rm d}z {\rm d}\varphi \label{eq:GalDist}
\end{eqnarray}
where $R_0 = 4.8 \, {\rm kpc}$ and $h_z = 1 \, {\rm kpc}$ \cite{GalacticDist}. The spin axes of 
black holes are assumed to be randomly directed.
It is rather difficult to estimate the absolute amplitudes of ringdown waves. 
Flanagan and Hughes have given an estimate on a 
fractional energy of the black hole masses radiated as gravitational waves to be 
about $3 \%$ \cite{Flanagan}, and we use this value for determination of the 
signal amplitudes. The injection point $t_0$
 is randomly selected in time of the TAMA300 observation periods. 
Once $t_0$ is given, 
the sky position of the source in the horizontal coordinates $(\theta, \phi)$ is 
obtained from the source position in the $(R, z, \varphi)$ coordinates, and then
the antenna pattern function in the direction $(\theta, \phi)$ is calculated.
The amplitude of the signal at the detector is determined from the distance to the
source, the antenna pattern function, 
and the angular distribution of the gravitational radiation determined
by the spin-weighted spheroidal harmonics ${}_{-2}S_{22}(\cos i, a)$, where $i$
is the inclination angle of the black hole axis seen from the observer.
The ringdown parameters $(f_c, Q)$ of the simulated signals are uniformly chosen 
in ranges $f_c = 100 \sim 2500 \, {\rm Hz}$ and $Q = 2 \sim 33.3$
(FIG. \ref{fig:parameter-range}, \ref{fig:templates}). The range of $f_c$ is 
determined considering the observational band of TAMA300 (see FIG. \ref{fig:sp}).
The range of $Q$ corresponds to the Kerr parameter $a = 0 \sim 0.998$.
The distribution of black hole masses (mass function) of the 
simulated ringdown signals results in
$\propto M^{-2}$, since $f_c$ is uniformly selected and $M$ is 
inversely proportional to $f_c$.

\begin{figure}
\begin{center}
\includegraphics[scale=0.31]{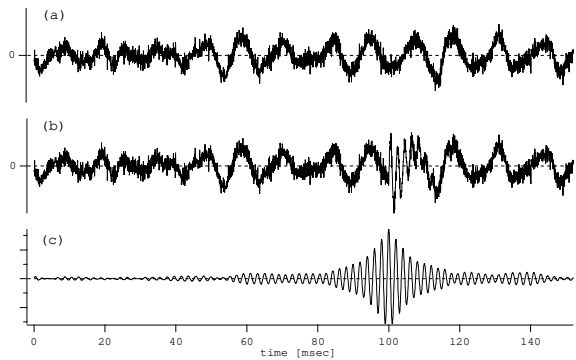}
\end{center}
\caption{Schematic view of software signal injection into real data (in time domain): (a) strain data $n(t)$ from the interferometer, (b) the strain data $n(t)$ and a simulated signal $s(t)$ with waveform parameters $f_c = 500 \, {\rm [Hz]}, Q = 20$ injected at $t_0 = 100 \, {\rm [msec]}$, (c) matched filter output $\left< n + s |  s \right>$.}
\label{fig:RDdemo}
\end{figure}

A simulated signal is superimposed on a data chunk of $52 \, {\rm sec}$ 
long at a randomly selected injection point $t_0$ with an initial phase 
(FIG. \ref{fig:RDdemo}), and the data chunk
is filtered with each template of the bank, and also with the 
waveform of the simulated signal itself.
The inputs for each event are the signal parameters $(f_c, Q)$, and the 
outputs are the matches for all of the filters used. Here the match for each template
 is given by equation (\ref{eq:inner0}) using the value of the filter output $\rho$ at
the local maximum found just around the injection point $t_0$. The SNR of an event 
is defined as the greatest in the matches, and the signal parameters
are determined as those of the {\it best-matched} template. The TAMA300 data
used here is obtained from 2001/Sep/17 to 2001/Sep/20 in TAMA Data Taking 6 (DT6),
and from 2003/Feb/13 to 2003/Feb/17 in DT8.
 For each data chunk, $200$ simulated signals are examined individually, 
and these procedures are carried out for all
the data chunks. The noise power spectrum of the detector outputs $S_n(|f|)$ required
in the calculation of the match is evaluated
for each data chunk with the moving average method using the preceding data chunks.
Note that we use the fixed template bank for all the data chunks, 
as described in section \ref{subsec:template}, and the amplitude of each template
is normalized so as to be $\left<h | h\right> = 1$ in equation (\ref{eq:inner0}).

This simulation method has some advantages in evaluating the detection 
efficiency or the detection probability for the Galactic events compared 
to the case with analytical calculations using a Galactic modeling and 
a typical detector sensitivity. Since we use the real 
observational data, we can include the time variation of detector's noise 
fluctuation and the 
sensitivity. Moreover, we also follow the antenna direction in the Galaxy 
at all times, thus we can properly take into account of the operational 
conditions of the detector 
to any Galactic events which would have occurred during the observation 
periods.

\section{Results}

\subsection{Detection Efficiency}
\begin{figure}
\begin{center}
\includegraphics[scale=0.47]{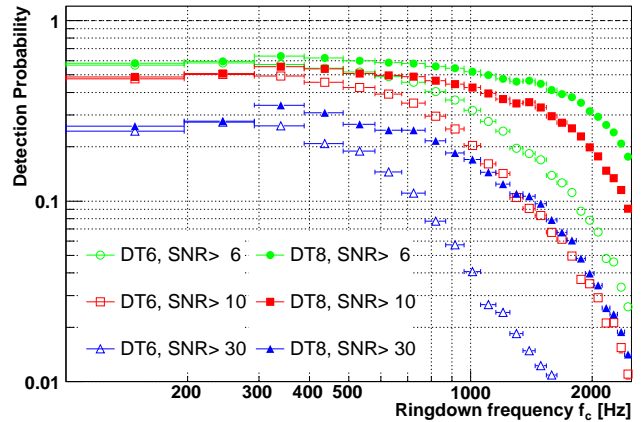}
\end{center}
\caption{Detection probability for Galactic events, for DT6 (open symbols) and DT8 (closed symbols). Circle: SNR $\ge 6$, Square: SNR $\ge 10$, Triangle: SNR $\ge 30$}
\label{fig:eff}
\end{figure}

By examining the SNRs of simulated ringdown events, we evaluate the detection 
efficiency $\epsilon$ for a given SNR threshold. Since we assume the sources are 
distributed in the Galaxy as equation (\ref{eq:GalDist}), 
this can be interpreted as the detection 
probability for the Galactic ringdown events. For example, if we detect
gravitational wave events $N_{\rm obs}$ in observation time $T$ 
and assume all of them are of Galactic
origin, the Galactic ringdown event rate is estimated as $N_{\rm obs}/(\epsilon T)$.
The detection probabilities in DT6 and DT8 as functions of the ringdown 
frequency $f_c$ are displayed in FIG. \ref{fig:eff} 
for different SNR thresholds, ${\rm SNR} \ge 6, \ge 10,$ and $\ge 30$. 
All the conditions which affect to the detection probabilities such as the 
template spacings, noise fluctuations and the position of the observer are properly 
taken into account. The detection 
probabilities are significantly improved in DT8, in particular for higher $f_c$ 
events. 
This is attributed to power-recycling implemented after 
DT6 to increase the input laser power in the cavities of the interferometer and 
reduce the shot noise which is dominant in higher frequencies. 
We can detect about $\epsilon = 50 \%$ of the Galactic ringdown events of
frequencies $f_c \le 600 \, {\rm Hz}$ (corresponding black hole mass 
$\ge 20 \, {\rm M_{\odot}}$) under the criterion ${\rm SNR} \ge 10$.

We note that these results depend on the evaluation of absolute amplitudes
of the ringdown waves at the sources. We assume that the fractional mass loss of a
black hole by radiating gravitational waves is $3 \%$ according to Flanagan and 
Hughes, and all of the energies are carried by 
the waves of the least damped mode of the QNMs. 

\subsection{Matched Filtering and Parameter Determination}
Next we consider matches between waveforms and templates, and 
accuracies in determination of the waveform parameters. 
It is expected that these are primarily limited by the template spacings.
We recall here the definition of the match between a template $h$ and 
a signal $s$ injected in detector output $n$:
\begin{eqnarray}
\rho = \left<o | h \right> \propto \int df \,
{\tilde{o}(f) \tilde{h}^*(f) \over S_n(|f|)} 
\label{eq:inner2}
\end{eqnarray}
where $\tilde{o}(f) = \tilde{n}(f) + \tilde{s}(f)$.
For each of our simulated signals, we can define three points in the parameter 
space, those are (1) a {\it priori} template with the same parameter values of 
the simulated signal, $\mathbb{S}: (f_c^S, Q^S; \rho^S = \left< o | \mathbb{S} \right>)$, 
(2) the nearest to $\mathbb{S}$
in the template bank, $\mathbb{N}: (f_c^N, Q^N; \rho^N = \left< o | \mathbb{N} \right>)$, 
and (3) the 
best-matched template, $\mathbb{M}: (f_c^M, Q^M; \rho^M = \left< o | \mathbb{M} \right>)$, 
which is the maximum of the match in the bank and gives the SNR for the signal.
The template we can single out from the bank is $\mathbb{M}$, not the nearest
 $\mathbb{N}$, but the implicit assumption in matched filtering is that the two 
templates $\mathbb{N, M}$ are identical and the parameters 
$f_c^N = f_c^M, Q^N = Q^M$ are the best estimates for $\mathbb{S}$.
Note that the metric of the parameter space which defines the distances between
the two templates depends on the noise power spectrum $S_n(|f|)$.
In the template construction we employ in this work, the metric of the parameter 
space is given on the assumption of white 
noise, $S_n(|f|) = {\rm constant}$.
 The template spacing is carried out in order 
that for any signals of waveform parameters within the search range, we can find at 
least one template such that the match is greater than the minimal 
match $MM$, i.e. $\rho^M/\rho^S \ge MM$. Thus if the noise power spectrum is colored, 
the definition of the distance is changed and hence the maximum of the match 
$\rho^M/\rho^S$ could be smaller than $MM$.
Nakano {\it et al.} has investigated the matches between
the templates and signals with randomly selected waveform parameters, 
$\left< s | \mathbb{N} \right>$,
by direct integration of equation (\ref{eq:inner2}) using an analytical fit 
of the TAMA300 typical noise curve (no lines, no fine structures), 
and showed that the fraction of signals which give $\rho^M/\rho^S < MM$ 
is only $0.88 \%$ (\cite{Nakano2}, see also section \ref{subsec:template}). 
Therefore we use the {\it fixed} template bank 
in the matched filtering analysis using TAMA300 data, for all the data chunks 
at different detector conditions.

\begin{figure}
\begin{center}
\includegraphics[scale=0.47]{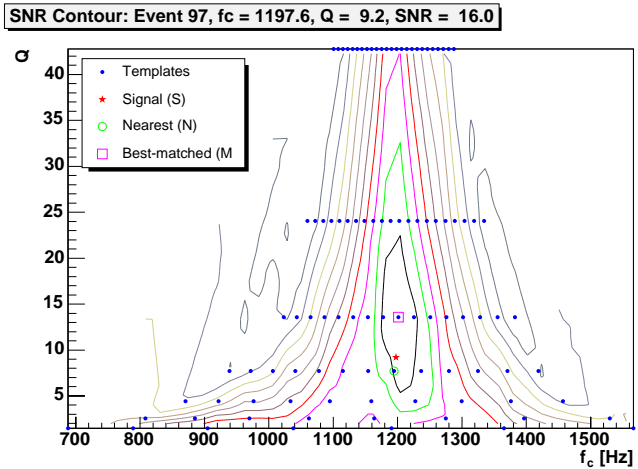}
\end{center}
\caption{An example of contour map of SNRs in the parameter space. 
The star is the simulated signal $\mathbb{S}$, which falls at 
$(f_c = 1197.6 {\rm Hz}, Q = 9.6)$, and the solid points represent the 
templates around $\mathbb{S}$. The circle indicates the nearest template $\mathbb{N}$, 
and the square shows the best-matched template $\mathbb{M}$ which gives the maximum 
of SNR. In most of events the two templates $\mathbb{N}$ and $\mathbb{M}$ 
coincide, but differ 
in this sample. This is due to an additional ``curvature'' of the parameter space 
introduced by the noise power spectrum $S_n(|f|)$, and contamination of the signal 
by non-stationary noises.}
\label{fig:contour}
\end{figure}

However, the noise power spectrum of a real detector $S_n(|f|)$ has many lines and fine 
structures, and changes its shape with time. 
Moreover, the detector noises are not stationary and
signals would be contaminated by burst-like noises. Thus it is not clear
whether the criterion $\rho^M/\rho^S \ge MM$ is still fulfilled in a practical search
using the {\it fixed} template bank. An alternative method is use of 
{\it adaptive} template banks, determined from the actual noise power spectrum 
and metric of the parameter space, as performed in neutron star 
binary search \cite{TakahashiChirp}. Our approach makes the data analysis codes
simpler, but the effect of noise fluctuations must be considered carefully.
Furthermore, when the noise power spectrum $S_n(|f|)$ 
is deformed and the signal $\tilde{s}(f)$ in equation (\ref{eq:inner2}) 
falls in a noisy frequency region, the match is degraded and
the template $\mathbb{M}$ selected from the bank does not necessarily coincide 
with the nearest template $\mathbb{N}$ (see FIG. \ref{fig:contour}).
In this case, the errors in parameter determination could be larger than those 
limited by the template spacings. 

\subsubsection{Finding Best-Matched Template}
\begin{figure}
\begin{center}
\includegraphics[scale=0.45]{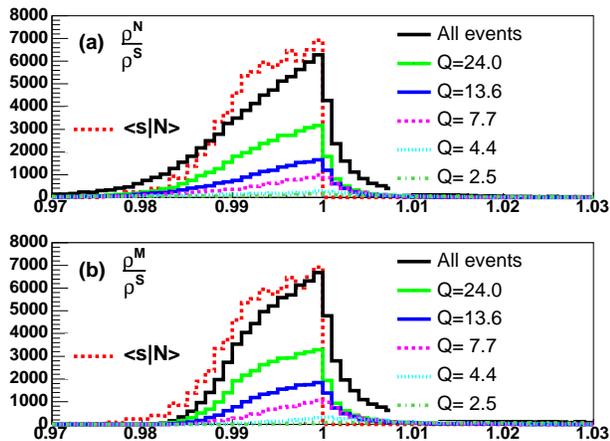}
\end{center}
\caption{Distributions of ratios of the SNRs, 
$\rho^N/\rho^S = \left< o | \mathbb{N} \right>/\left< o | \mathbb{S} \right>$ 
and $\rho^M/\rho^S = \left< o | \mathbb{M} \right>/\left< o | \mathbb{S} \right>$ (see text for definitions).
 The broken line is distribution of $\left< s | \mathbb{N} \right>$, where $s$ is 
a randomly generated ringdown waveform, 
obtained by direct integration of equation (\ref{eq:inner2}) using a
typical TAMA300 noise curve \cite{Nakano2}.}
\label{fig:snrloss}
\end{figure}

To investigate these effects, we examine the ratios 
$\rho^N/\rho^S = \left< o | \mathbb{N} \right>/\left< o |  \mathbb{S} \right>$ and 
$\rho^M/\rho^S = \left< o | \mathbb{M} \right>/\left< o | \mathbb{S} \right>$, those are measures of
the utility of the template bank.
If the noise power spectrum is white ($S_n(|f|) = {\rm const.}$) and there are no 
non-stationary noises, the maximum value of $\rho^N/\rho^S$ is unity
and its distribution has a cut-off
at the minimal match $MM = 0.98$ defined in the template construction.
The distribution of $\rho^N/\rho^S$ for the TAMA300 data-injected signals
are shown in FIG. \ref{fig:snrloss} (a).
For a reference the distribution of $\left< s | \mathbb{N} \right>$ 
calculated by using an analytical fit of the TAMA300 typical noise curve \cite{Nakano2}
is also displayed. The tail of the distribution 
of $\rho^N/\rho^S$ in the lower side looks in agreement to that expected 
from the template design, $\left< s | \mathbb{N} \right>$. 
The fraction of events with $\rho^N/\rho^S < MM$ is
$3.3 \%$, which is larger than $0.88 \%$ for $\left< s | \mathbb{N} \right>$ because of
the noise effects.
We note that about $10 \%$ of simulated signals have the fractional match 
$\rho^N/\rho^S$ greater than $1$ (many of them are observed in lower $Q$ 
signals), and the mean of $\rho^N/\rho^S$
is $0.994$ , slightly larger than that of $\left< s | \mathbb{N} \right>$, $0.993$.
This is related to contaminations of the signals by the noise $n$, and
the estimated noise power spectrum $S_n(|f|)$ used in matched filtering.
If the true noise power spectrum at the time when 
the data $n$ was obtained differs from the estimated spectrum 
$S_n(|f|)$, $\rho^N/\rho^S > 1$ is possible since each template is normalized 
as $\left< h | h \right> = 1$ by using the estimated spectrum $S_n(|f|)$. Therefore in a realistic
gravitational wave search there are 
inevitable ambiguities in evaluation of matches (or SNRs) and their losses.

In FIG. \ref{fig:snrloss} (b), the distribution of the ratio $\rho^M/\rho^S$
is shown. The mean of $\rho^M/\rho^S$ is $0.998$, and the fractions of
events of $\rho^M/\rho^S < MM$ and $\rho^M/\rho^S > 1$ are $0.0012 \%$ and $24\%$, 
respectively. This indicates that our template bank fulfills the requirement of the
minimal match $> MM$ even in practical search, and
also it is difficult to estimate the 
``true'' value of the match, $\left< s | \mathbb{S} \right>$ or $\left< s | \mathbb{N} \right>$,  
within accuracies of a few \%. The occurrence probability of 
$\mathbb{N} \ne \mathbb{M}$ is about $30 \%$ (see FIG. \ref{fig:singleout}). This causes increases of waveform parameter
determination errors.

\subsubsection{Waveform Parameter Errors}
\begin{figure}
\begin{center}
\includegraphics[scale=0.7]{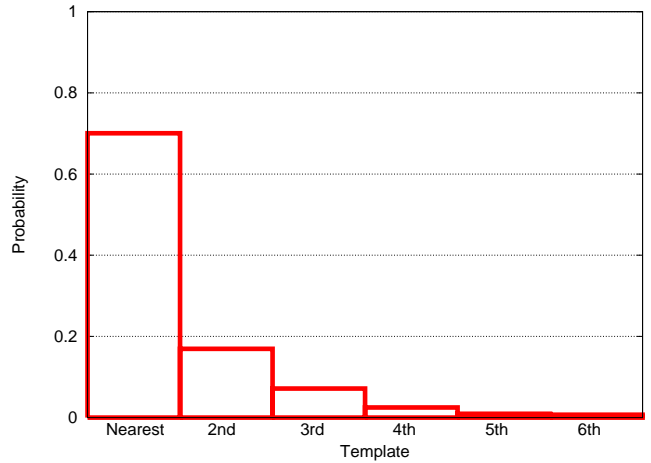}
\end{center}
\caption{Distribution of the templates which give the maximum of SNRs in the bank.}
\label{fig:singleout}
\end{figure}

The accuracies in determination of the waveform parameters $(f_c, Q)$ 
are evaluated by comparing those of the simulated signal $\mathbb{S}$ and of the 
best-matched template $\mathbb{M}$. 
The root mean squares (RMSs) of the errors $\left( \Delta f_c/f_c \right)_{\rm RMS} \equiv \left( (f_c^M - f_c^S)/f_c^S \right)_{\rm RMS}$ 
and $\left( \Delta Q/Q \right)_{\rm RMS} \equiv \left( (Q^M - Q^S)/Q^S \right)_{\rm RMS}$ 
are calculated as $1.3 \%$ and $22 \%$, 
respectively. These are comparable to
RMSs of $\left( \Delta f_c/f_c \right)_{\rm RMS} = 1.22 \%$ and 
$\left( \Delta Q/Q \right)_{\rm RMS} = 16.3 \%$ 
estimated from the template spacings, but somewhat broadened.
This is because for not all the events the template
$\mathbb{M}$ coincides with the nearest $\mathbb{N}$, 
as seen in FIG \ref{fig:singleout}.
In practice, $\left( \Delta f_c/f_c \right)_{\rm RMS}$ 
and $\left( \Delta Q/Q \right)_{\rm RMS}$ 
depend on $Q$, as the template distances are (see FIG. \ref{fig:templates}). 
The values of $\left( \Delta f_c/f_c \right)_{\rm RMS}$ 
and $\left( \Delta Q/Q \right)_{\rm RMS}$
are listed in TABLE \ref{table:parameter-errors}.

\section{Discussion}\label{secDisc}
\subsection{Black Hole Spectroscopy}
The distances between the ringdown filters 
determined from the requirement to the minimal match $MM$ limit the waveform parameter
resolutions. The resolutions for the ringdown parameters
in matched filtering analysis with the template bank of $MM = 0.98$ 
using TAMA300 data are $\left( \Delta f_c/f_c \right)_{\rm RMS} = 1.3 \%$ and 
$\left( \Delta Q/Q \right)_{\rm RMS} = 22\%$ as described in the 
previous section. The question here
is how accurately we can determine the source parameters, the mass $M$ and the Kerr
parameter $a$, from the results of matched filtering analysis. 
By using the Echeverria's expressions (\ref{eq:fc}) and (\ref{eq:Q}),
we found $\left( \Delta M/M \right)_{\rm RMS} = 10 \sim 20 \%$ 
for low $Q$ signals, and a few $\%$ for higher
$Q$ ringdowns. The error in the Kerr parameter determination is about
$\left( \Delta a/a \right)_{\rm RMS} = 40 \sim 60 \%$. The calculated values of 
$\left( \Delta M/M \right)_{\rm RMS}$ and $\left( \Delta a/a \right)_{\rm RMS}$ 
are listed in TABLE \ref{table:parameter-errors}. 

\begin{figure}
\begin{center}
\includegraphics[scale=0.68]{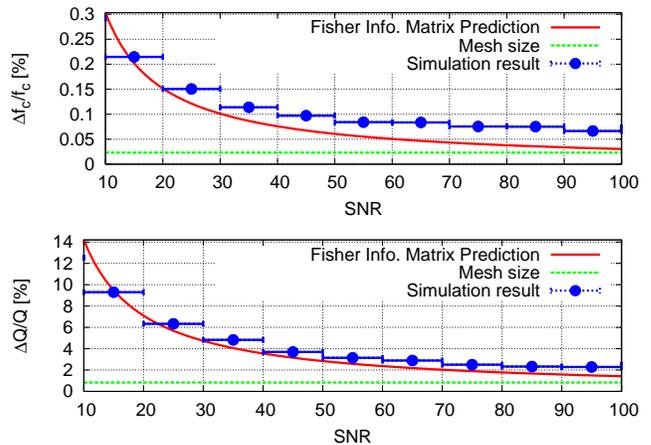}
\end{center}
\caption{Waveform parameter resolutions in the case of a finely meshed template bank
(closed circle). The simulation is carried out for a fixed waveform with parameters 
$f_c = 580 {\rm Hz}$ and $Q = 24$. 
The templates are uniformly placed around the signal with separations
of $\Delta f_c = 0.13 {\rm Hz} \, (\Delta f_c/f_c = 0.023 \%)$ 
and $\Delta Q = 0.2 \, (\Delta Q/Q = 0.83 \%)$ (broken line). The predictions from
the Fisher information matrix, which are expected in cases of Gaussian noises 
are also shown (solid line).}
\label{fig:snr_sigma}
\end{figure}

Obviously these results depend on the template 
design, i.e. template spacings. It is expected that the parameter resolution 
can be improved as template distances become closer, and saturates 
when the mismatches between signals and neighboring templates approach 
to intrinsic ambiguities limited by the noise fluctuations. 
In case of Gaussian noise, the intrinsic parameter
estimation errors can be evaluated by using the Fisher information matrix 
\cite{Finn,CutlerFlanagan}. For example, for 
ringdown signals of $(f_c = 580 {\rm Hz}, Q = 24)$,
these are calculated as functions of SNRs, as
$\left( \Delta f_c/f_c \right)_{\rm RMS} = 0.30\%$ 
and $\left( \Delta Q/Q \right)_{\rm RMS} = 14 \%$ at ${\rm SNR} = 10$
and inversely proportional to SNRs (see Appendix B. in \cite{Nakano2}).
To confirm this, we performed matched filtering simulations 
for fixed parameter ringdown signals
at $f_c = 580 \, {\rm Hz}, Q = 24.0$ ($M = 50 M_{\odot}, a = 0.996$),
 with very fine-meshed templates
uniformly distributed around the signal. 
We found $\left( \Delta f_c/f_c \right)_{\rm RMS} = 0.21\%$ 
and $\left( \Delta Q/Q \right)_{\rm RMS} = 9.3 \%$ for signals of
 ${\rm SNR} = 10 \sim 20$ (FIG \ref{fig:snr_sigma}).
The parameter estimation errors are in agreement with the predictions 
of the Fisher information matrix, and small 
excursions from them are considered due to non-Gaussian component of the TAMA300 data.
With these results, the black hole parameter estimation errors are evaluated as
$\left( \Delta M/M \right)_{\rm RMS} \sim 0.9 \%, \left( \Delta a/a \right)_{\rm RMS} \sim 24 \%$ for ${\rm SNR} \sim 10$, and
$\left( \Delta M/M \right)_{\rm RMS} \sim 0.3 \%, \left( \Delta a/a \right)_{\rm RMS} \sim 7 \%$ for ${\rm SNR} \sim 50$.
No further improvements are expected even in a case of more finer mesh,
since these limits are from the intrinsic ambiguities in matches due to
the noise fluctuations. 
(Note that since the accuracies in the conversions from
$(f_c, Q)$ to $(M, a)$ are only $\sim 5 \%$,
we need more accurate formulae.)

When we find event candidates of ringdown signals, 
we can determine the black hole parameters within the accuracies described above 
by setting a fine-meshed template bank locally at each
of the event candidates. There is no need to cover whole parameter space with the 
template bank of such closer distances, the hierarchical search strategy copes with 
computational costs \cite{Mohanty}.
The resolutions for the black hole parameters obtained here are remarkably good,
compared to those estimated dynamically for known black hole candidates in 
X-ray binaries or radio sources, typically $10 \sim 100 \%$ \cite{Ziolkowski}. 
Some X-ray binaries as GRS1915+105 eject superluminal jet, and rotating black holes
are considered as the central engine \cite{Greiner}.
Precise determination of the Kerr parameter $a$ of black holes by observing 
gravitational waves will give strong
constrains on jet formation mechanisms \cite{Koide,Villiers,Tomimatsu}.
Thus gravitational wave observations provide not only a probe for 
black hole detections, but also a promising clue to establish 
black hole ``spectroscopy''. 

\begin{table}
\begin{center}
\begin{tabular}{|c||cc|cc|} \hline
$Q^M$ & $\left(\Delta f_c/f_c\right)_{\rm RMS}$ & $\left(\Delta Q/Q\right)_{\rm RMS}$ & $\left(\Delta M/M\right)_{\rm RMS}$ & $\left(\Delta a/a\right)_{\rm RMS}$ \\ \hline
All & 1.3 (1.2) & 22 (16) & &  \\
2.55 & 8.1 (2.6) & 22 (16) & 22 (12) & 64 (35) \\
4.41 & 4.0 (1.6) & 24 (16) & 13 (6.6) & 41(35) \\
7.70 & 1.6 (1.0) & 21 (16) & 6.8 (3.9) & 39 (36) \\
13.6 & 0.77 (0.58) & 19 (16) & 3.1 (2.4) & 40 (36) \\
24.0 & 0.39 (0.33) & 19 (17) & 1.9 (1.6) & 41 (37) \\ \hline
\end{tabular}
\end{center}
\caption{RMS errors in estimations of ringdown - black hole parameters (in $\%$) 
by using the template bank of $MM = 0.98$.
The quantities are defined as 
$\Delta X/X \equiv (X^M - X^S)/X^S, \, (X = f_c, Q, M, a)$.
For comparison, analytic 
estimations from the template distances, which should be obtained in the absence 
of noise fluctuations, are also shown (parenthesized). 
The parameter conversions $(f_c, Q) \leftrightarrow (M, a)$ are 
based on the Echeverria's expressions in equations (\ref{eq:fc}) and (\ref{eq:Q}).}
\label{table:parameter-errors}
\end{table}

\section{Conclusions}\label{secConc}
Black hole ringdowns are promising source of gravitational radiation, and
the matched filtering technique is suitable to extract ringdown signals 
from outputs of gravitational wave detectors. In this paper, we studied
template constructions, detection efficiencies and waveform parameter 
estimations in matched filter searches using TAMA300 data. In template
construction, we employed a simple and efficient tiling method 
proposed by Nakano {\it et al.} \cite{Nakano,Nakano2}.
The template bank constructed with this method assures the {\it minimal match}
criterion for any ringdown waveforms in the cases of 
a flat noise power spectrum $S_n(|f|) = {\rm const.}$ and a colored spectrum
fitted to TAMA300 typical noise curve \cite{Nakano,Nakano2}. The errors in
waveform parameter estimation are determined by the template separations
hence the choice of $MM$.
In gravitational
wave searches using real observational data, however, it is not clear
whether the minimal match criterion is fulfilled since non-stationary noises
contaminate signals and the time-varying noise power spectra 
change {\it matches}
or {\it distances} between signals and the templates. In a such case,
the {\it nearest} template and the {\it best-matched} template for a signal
do not necessarily coincide and parameter estimation errors increases.

To investigate these
effects, we performed matched filtering analysis for artificial ringdown
signals injected into TAMA300 data with Monte-Carlo technique. We consider
black holes with masses $10 \sim 200 M_{\odot}$ distributed in the Galaxy
as ``realistic'' search targets. With TAMA DT8 sensitivity and using our 
template bank, the detection probability is evaluated as $\sim 50 \%$ 
for Galactic ringdown events of $f_c < 600 {\rm Hz}$, which correspond to
black holes with masses $\ge 20 M_{\odot}$, assuming the fractional black 
holes mass $\sim 3 \%$ radiated as gravitational waves. 
We confirmed that the loss of SNR is less than $2 \%$ in most cases
for ringdown signals embedded in TAMA300 data, therefore we can use the
template bank in gravitational wave searches using real observational data. 
We also showed that it is difficult to evaluate 
matches $\rho$ or SNRs within an accuracy of a few $\%$ because of 
the signal contaminations and the difficulties in estimation of 
``true'' noise power spectrum at given observation time.
The RMS errors in the waveform parameter estimations are examined
by comparing the simulated signals and the templates singled out from the bank.
We obtained $\left( \Delta f_c/f_c \right)_{\rm RMS} = 1.3 \%$ 
and $\left( \Delta Q/Q \right)_{\rm RMS} = 22 \%$, which are comparable with the 
expectations from the template spacings, 
$\Delta f_c/f_c = 1.22 \%$ and 
$\Delta Q/Q = 16.3 \%$, respectively.
The informations obtained in this work as the detection probability 
and the parameter estimation errors will be used in statistical and 
astrophysical interpretation 
of results from the TAMA300 full-scale event search, which is under progress.
The key in the event search is to develop efficient 
vetoing techniques to distinguish real signals and spurious triggers due to 
impulsive noises \cite{TsunesadaCQG1}.

Our interest extends to how accurately we can determine the waveform parameters
hence the black hole masses and the angular momenta by using an 
extremely fine-meshed template bank. For ringdown signals of 
$f_c = 580 \, {\rm Hz}, Q = 24.0$, the ultimate resolutions are evaluated 
as $\left( \Delta f_c/f_c \right)_{\rm RMS} = 0.21\%$ 
and $\left( \Delta Q/Q \right)_{\rm RMS} = 9.3 \%$ for signals of
 ${\rm SNR} = 10 \sim 20$, which are
consistent with the predictions of the Fisher information matrix.
The errors in black hole parameter estimations are calculated 
as $\left( \Delta M/M \right)_{\rm RMS} \sim 0.9 \%, \left( \Delta a/a \right)_{\rm RMS} \sim 24 \%$ for ${\rm SNR} \sim 10$, and
$\left( \Delta M/M \right)_{\rm RMS} \sim 0.3 \%, \left( \Delta a/a \right)_{\rm RMS}  \sim 7 \%$ for ${\rm SNR} \sim 50$.
In a practical search, we will employ the hierarchical search strategy,  
performing the first event search with an ``ordinary'' template bank, 
of minimal match $MM = 0.98$ for example, and the second search 
for event candidates with a locally fine-meshed template bank.
The remarkably good resolutions for $(M, a)$ are 
heartening to understand gravitational physics and astrophysics of black holes.

\begin{acknowledgments}
The authors wish to acknowledge to all the collaborators in TAMA project.
YT expresses thanks to T. Akutsu, K. Arai, M. Fujimoto, K. Hayama, S. Sato,
K. Somiya, and R. Takahashi for helpful comments and encouragements.
HN would like to thank H. Ishihara and K. Nakao for useful discussions.
This research is partially 
supported by a Grant-in-Aid for Creative Basic Research (09NP0801), 
for Scientific 
Research on Priority Areas (415), and for Scientific Research No. 16540251
of the Ministry of Education, 
Culture, Sports, Science and Technology.
HN is supported by the Japan Society for the Promotion 
of Science for Young Scientists (No. 5919).
\end{acknowledgments}

\appendix
\section{Preceding Waveforms}
\begin{figure}
\begin{center}
\includegraphics[scale=0.68]{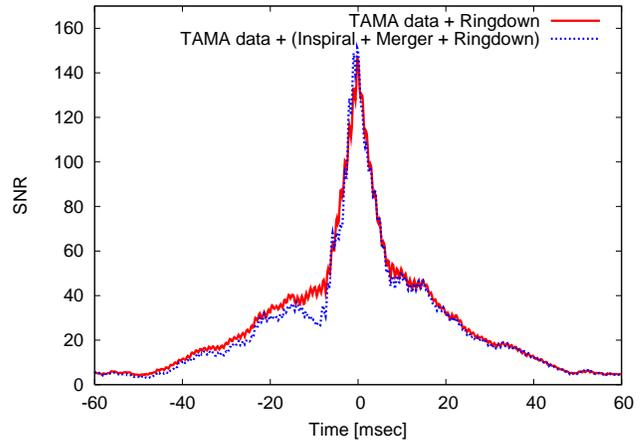}
\end{center}
\caption{An example of ringdown filter outputs in time series 
for a ringdown signal (solid) and for a composite waveform 
(broken) injected in a TAMA300 data chunk.}
\label{fig:composite}
\end{figure}

So far we have assumed that there are no preceding waves before 
the ringdown signal arrives. If the ringdown is initiated by a binary coalescence, 
however, we would receive gravitational waves of characteristic waveforms 
successively, as the chirp signal at first, the waves emitted in the merger 
phase, followed by the ringdown signal in the last stage.
To investigate effects of the preceding waves,
we performed a test of matched filtering for a such composite waveform 
injected in TAMA300 data. The aim of this is not to generate
realistic gravitational waveform from a binary coalescence, nor to use 
composite waveforms as templates, but to examine whether we can extract 
ringdown signals alone from the composite waveform embedded in detector 
outputs by using our ringdown filters. Here we considered binary coalescence 
of masses $25 M_{\odot} - 25 M_{\odot}$ to form a black hole of $50 M_{\odot}$ 
with a Kerr parameter $a = 0.98$ ($f_c = 520 {\rm Hz}, Q = 12$).
In this case, the cut-off frequency of the inspiral phase
is below $100 {\rm Hz}$, therefore the contribution of the chirp signal is 
small. Although the waveforms in the merger phase are quite uncertain,
we presume sine-Gaussian waveform, as commonly used in the burst search 
studies, of the central frequency $\sim 300 {\rm Hz}$ with amplitude of 
twice of the ringdown signal and a phase to be connected analytically with 
the following ringdown. We found that (i) the maximum of SNR is found 
at the template nearest to the simulated signal, (ii) the peak of the 
filter output in time series is found at which ringdown begins, 
and (iii) the value of SNR is consistent
within $1 \%$ compared to that in the case of no preceding waveforms injected 
(FIG. \ref{fig:composite}).  This is just a preliminary inspection 
on effects of the preceding waveforms in a ringdown search, 
and we need to know more physics of the merger phase for
quantitative examinations.

\bibstyle{prd}
\bibliography{ringdown}

\end{document}